# Intelligent Control of Transportation Flow in *Physarum* Networks

Bingyang Han[*], Luolan Chen and Tieyan Si[#]

School of Physics, Harbin Institute of Technology, Harbin 150001, China

[*] **E-mail:** hanbingyang2023@163.com
[#] **E-mail:** tieyansi@hit.edu.cn

**Abstract:**

The *Physarum* network expands or retracts in response to environmental stimuli, demonstrating an intelligent adaptive capability to locate optimal paths for nutrient transport. The underlying physical mechanism governing this intelligence behavior remains an unresolved problem in biological physics. Unlike the unidirectional flow typical of urban traffic networks, cytoplasmic flow within the *Physarum* network exhibits periodic oscillations modulated by biological repellents and attractants. In this study, we investigate how local flows within the network's branch channels interact to collectively govern the global oscillatory dynamics. We find that the measured flow fluxes at intersection nodes obey Kirchhoff's current law. Phase differences exist among the flows in different branches. At the microscopic scale, flow distribution exhibits only brief periods of traffic congestion, which are resolved by the oscillatory flows. By mapping the flow flux vectors onto the magnetic moment vector of spin ice model, we demonstrate that the flow vectors strictly obey the ice-rule of vertex models in statistical physics. Notably, the three branches converging at a Y-shaped node never become blocked simultaneously, thereby preventing traffic congestion and ensuring efficient transmission of nutrients and signals. This intelligent flow control phenomenon offers novel insights for addressing traffic congestion and advances our understanding of frustrated quantum magnetism.

## 1 Introduction

Unlike conventional polymer network structures in soft condensed matter[1], the *Physarum* network is an active, adaptive system. It exhibits both protozoan motility and highly specific propagules at different stages of their life[2, 3]. Although its morphology structure and physiology share similarities with fungi[4, 5], but demonstrating emergent intelligence for searching and redistributing foods. This remarkable capability has inspired innovations in diverse fields, including biologically engineered systems[6],anticancer therapeutics, and agricultural[7, 8]. Furthermore, the *Physarum* network can retract redundant connections and dynamically reconfigure itself into an optimized architecture for efficient nutrient transport. This adaptive behavior has offered valuable insights for applications in ecology, urban planning, and transportation network design[9-12], provided a natural simulation for studying optimal transport theory in computer and information sciences[13-17].

The cytoplasmic flow within the *Physarum* network delivers oxygen and nutrients to all nodes, thereby sustaining cellular steady growth and homeostasis[18]. This flow is not unidirectional but generated by the network's self-organized, periodic contractions[19]. Termed the "venous shuttle flow", it rhythmically transports nutrients back ang forth within the veins[20]. The fundamental physical and biological mechanism governing the emergence of this shuttle flow remains an unresolved interdisciplinary problem.

The oscillatory dynamics and network growth are modulated by environmental chemical stimuli, as different nutrients and repellents elicit distinct chemotactic responses[21, 22]. At the molecular level, the cytoskeleton composed of interconnected actin fibers, provides the structural framework. Network growth and cytoplasmic streaming are driven by the periodic contraction and relaxation of actin-myosin interactions within the veins, a process powered by ATP hydrolysis and modulated by $Ca^{2+}$ ions [23-26]. Consequently, growth can be suppressed in complex environments containing both high nutrient concentrations (e.g., glucose as an attractant) and repellents (e.g., NaCl) [27].

To elucidate the underlying physical mechanism of the venous shuttle flow, we quantitatively measured the protoplasmic flows in multiple branches converging at the same node. This investigation aims to understand how cytoplasmic flows within local veins cooperate to generate collective oscillations and prevent traffic congestion.

## 2 Results

### 2.1 The branched flow fluxes at the Y-shape intersection node

The *Physarum* network consists of numerous interconnected Y-shaped nodes, where three tubular veins converge. Fluorescence microscopy revealed that intersections of more than three veins at a single node are exceedingly rare. We measured the flow flux within veins at Y-shaped nodes under four different environmental conditions: (1)100mM glucose, (2)200mM glucose, (3)100mM NaCl, and (4)a blank control. Corresponding microscopic images of the vein structures in this four solutions are presented in Fig 1. The lumens of the veins are filled with a viscous fluid containing active particulates[18], including ATPase particles involved in energy metabolism, $Ca^{2+}$ vesicles responsible for signal transduction[28], and actin-myosin complexes that mechanically regulate the cytoskeleton[29]. The high concentration of these particulates results in a low-Raynolds-number flow regime[30, 31].

To quantify oscillations in the cytoplasmic flow, one type of these active particulates was selected as a tracer and tracked in recorded videos. The resulting particle trajectories confirmed the laminar nature of the flow within the tubular veins. Subsequently, the volumetric flow rate, Q, through a vein was calculated as the product of the time-averaged flow velocity and the vessel's cross-sectional area, as summarized in the following equation,

$$Q = S \cdot v = \pi r^2 \cdot \sum_{i=1}^{n} v_i \tag{1}$$

According to the conservation law of total mass, the flow flux should obey the same Kirchhoff's first law as that in circuit electric current theory.

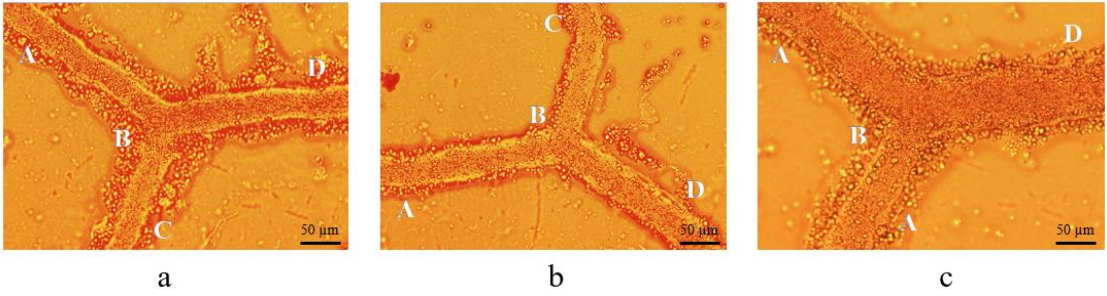

Figure 1 The microscope image of the branch veins at the Y-shaped intersection node under a 20× objective lens of an upright fluorescence microscope in three different solutions. (a) in 100mM Glucose solution; (b) in 100mM NaCl solution; (c) in 200mM Glucose.

$$\sum_{j} Q_j = 0 \tag{2}$$

The flow velocity of the protoplasm was measured by the hydrological buoy method[32]. The internal diameter of each vein was obtained from video recordings by measuring the lumen width. The reported flow velocity was calculated as the average over ten discrete time steps. The cytoplasmic flow fluxes within the three branches under each experimental condition are summarized in Table 1 (100mM Glucose), Table 2 (200mM Glucose), and Table 3 (100mM NaCl). Although the absolute flux values differ among the three solutions, they all satisfy Kirchhoff's first law (Eq. 1).

Table 1: The measured flow within the branches of the Y-shaped *Physarum* network in a 100mM Glucose environment.

| Vein | Average Speed(mm/s) | Diameter (mm) | Sectional area(mm$^2$) | Traffic (mm$^3$/s) | Direction |
|---|---|---|---|---|---|
| AB | 9.21×10$^{-2}$ | 3.00×10$^{-2}$ | 7.09×10$^{-4}$ | 6.53×10$^{-5}$ | Into |
| CB | 1.26×10$^{-2}$ | 3.79×10$^{-2}$ | 1.13×10$^{-3}$ | 1.42×10$^{-5}$ | Out |
| DB | 4.74×10$^{-2}$ | 3.71×10$^{-2}$ | 1.08×10$^{-3}$ | 5.11×10$^{-5}$ | Out |

Table 2: The measured flow within the branches of the Y-shaped *Physarum* network in a 100mM

NaCl environment.

| Vein | Average Speed(mm/s) | Diameter (mm) | Sectional area(mm$^2$) | Traffic (mm$^3$/s) | Direction |
|---|---|---|---|---|---|
| AB | $1.73\times10^{-2}$ | $4.67\times10^{-2}$ | $1.71\times10^{-3}$ | $2.96\times10^{-5}$ | Into |
| CB | $5.73\times10^{-3}$ | $4.33\times10^{-2}$ | $1.47\times10^{-3}$ | $8.43\times10^{-6}$ | Into |
| DB | $1.87\times10^{-2}$ | $5.12\times10^{-2}$ | $2.07\times10^{-3}$ | $3.88\times10^{-5}$ | Out |

Table 3: The measured flow within the branches of the Y-shaped *Physarum* network in a 200mM Glucose environment.

| Vein | Average Speed(mm/s) | Diameter (mm) | Sectional area(mm$^2$) | Traffic (mm$^3$/s) | Direction |
|---|---|---|---|---|---|
| AB | $9.26\times10^{-2}$ | $5.59\times10^{-2}$ | $2.46\times10^{-3}$ | $2.28\times10^{-4}$ | Out |
| CB | $4.70\times10^{-2}$ | $4.83\times10^{-2}$ | $1.83\times10^{-3}$ | $8.60\times10^{-4}$ | Out |
| DB | $9.52\times10^{-2}$ | $6.49\times10^{-2}$ | $3.31\times10^{-3}$ | $3.15\times10^{-3}$ | Into |

## 2.2 The branch flows at the H-shaped intersection nodes

The flow within branch veins of the *Physarum* network does not reverse direction simultaneously. Instead, they self-organized into a time serial ordering to reverse the local branch flow one after another. There always exists a jammed branch vein when the other two branches reverse their flow direction at the Y-shaped node. As shown in Fig 2, an H-shaped network consists of two Y-shaped nodes. In this configuration, a jammed vein at one node directly affects the traffic flow at the adjacent node. The jammed state is short-lived, representing an intermediate phase between two states of opposite flow.

As shown by the green arrows in upper panel of Fig 2b, the input flow from entrance

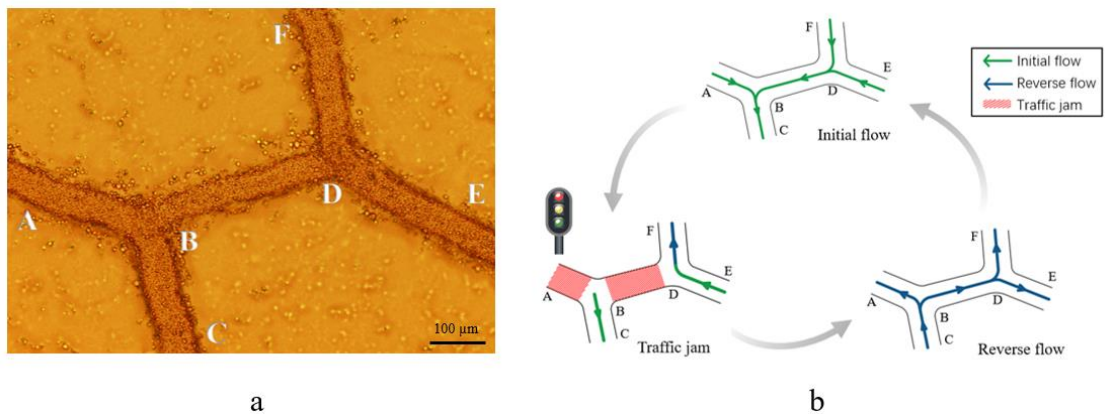

Figure 2 The branch veins at the H-shaped node. (a) Images taken under a 10x objective with an upright fluorescence microscope. (b) One periodicity of oscillating flows in *Physarum* networks.

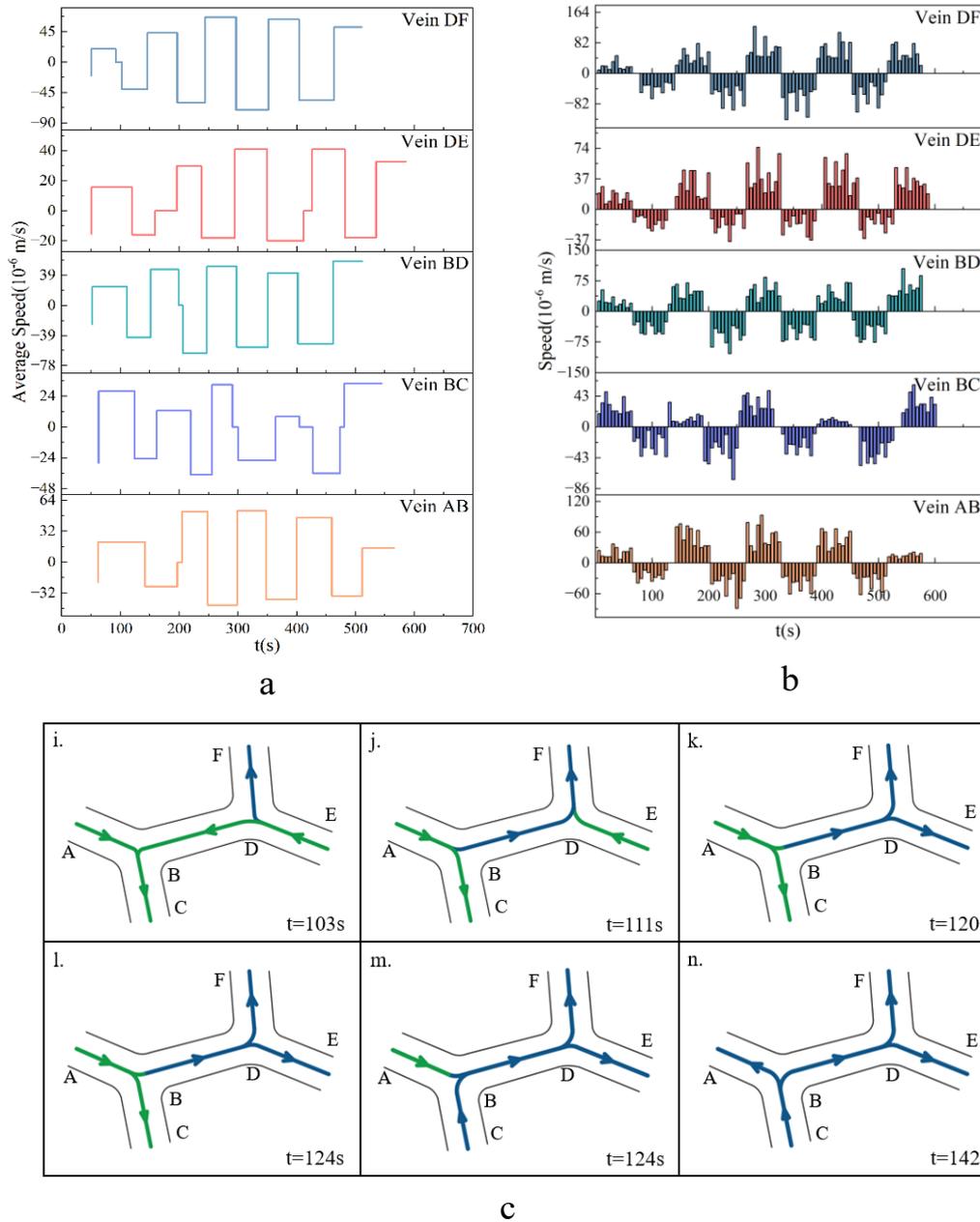

Figure 3. (a) The periodical oscillation of average velocity in all branches of the H-shaped *Physarum* network. (b) The periodical oscillation of simultaneous velocity within all branches. (c) The reversion time serials of velocity within the branches of the H-shaped network. i) the initial flow velocity configuration. j) BD branch reverses its velocity at 111s. k-l) The DE branch t reverses its velocity at 120s. m) The BC branch reverses its velocity at 124s. n) The AB branch reverses its velocity at 142s.

E and entrance F converge at node D, then flow to node B together with the input from entrance A. The flow velocity gradually decays until there is traffic congestion occurs along the branch AB and branch DB in the left panel of Fig 2b. After several seconds, the flow in branches FD and DE reverses direction first. This reversal displaces the jammed particles from branch BD, which subsequently triggers a flow reversal in all branches, as shown in the right panel of Figure 2b. The jammed branch AB is diluted by the flows out of the H-shaped node. The three input entrances (A, F and E), which fuse into one exit at C, periodically transform into three exits and one input entrance at C.

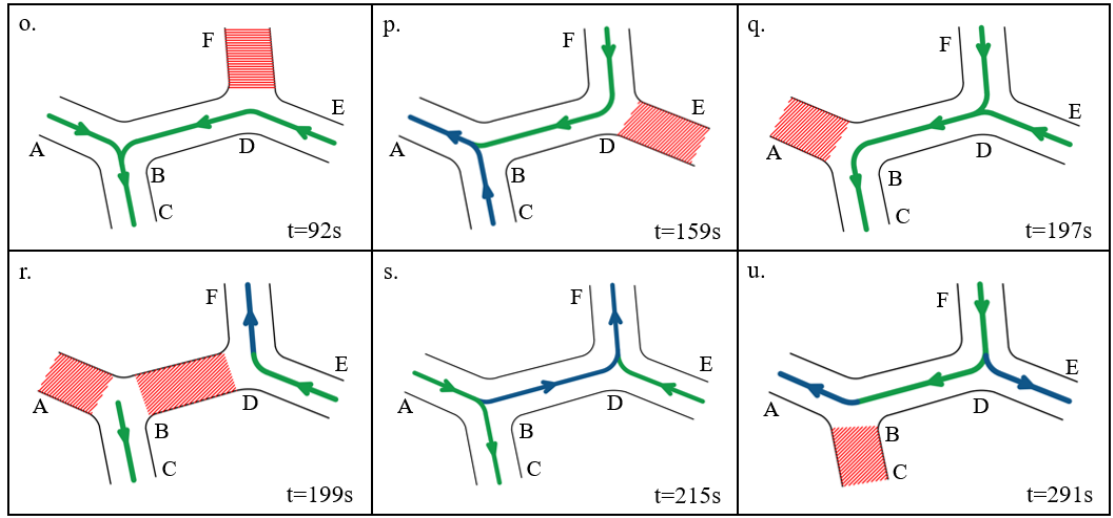

Figure 4: The time serial of jammed state within the five branches of the H-shaped network. (o) The FE is jammed at 92s. (p) The DE is jammed at 159s. (q) The AB is jammed at 197s. (r) The BD is jammed at 199s. (s) The BC is jammed at 215s. (u) The BC is jammed at 291s.

Periodic oscillations in branch flow velocity were quantified from video recordings obtained under microscopy, as summarized in Fig 3. Fig 3a presents the velocity averaged over ten time points for five branch veins. Fig 3b displays the corresponding instantaneous velocity, which exhibits clear periodic oscillations as further illustrated in Fig 3c. The observed sequence of flow reversals begins when vein BD reverses direction, redirecting flow region A toward F. This is followed by a reversal in the vein DE, which splits the flow within BD. The resulting flow deficit in BD is then compensated by a concurrent reversal in the vein BC; this reversal in BC also diverts part of its flow into the now-reversed vein BA. This coordinated reversal sequence repeats periodically across all branches.

The plateau regions of zero flow velocity correspond to the traffic congested state. As shown by the oscillatory velocity profiles in Fig 3a, the flows in the five branch veins do not reverse direction simultaneously. Instead, a clear phase difference exists among the oscillatory cycles of different veins. The H-shaped network consists of two Y-shaped nodes. At each Y-shaped node, the three connecting veins never become blocked concurrently. For example, when inflow from entrance F is obstructed, current is redirected from entrance E toward region B (Fig 4-o). Conversely, when entrance E is blocked, flow from entrance F is redirected to B. A similar regulatory pattern is observed at node B, as illustrated in Fig 4q-4u.

The *Physarum* network comprises numerous Y-shaped nodes with branch veins of different scales. Locally, a small region of the network can be modeled as a three-vertex system of interconnected Y-shaped nodes. A fundamental distinction from the classical spin ice model in statistical physics[33, 34] lies in the periodic reversal of flow vector orientations within the *Physarum* network. As shown in Fig 5a-I, the "two-in-one-out" and "one-in-two-out" flow configurations, which correspond to the ice-rule in spin ice model, are consistently observed at every Y-shaped node. During the flow reversal period, one of the three veins becomes transiently jammed (Fig 5a-II). In contrast, the "three-in and three-out" configuration depicted in Fig 5a-III, which violates the ice rule, was neither observed experimentally nor generated by simulations. In spin ice systems, such a configuration raises the total energy of the vertex lattice, which is disfavored in ground state. In the *Physarum* network, this configuration would induce traffic congestion; however, it is prevented by the network's inherent periodic flow reversal.

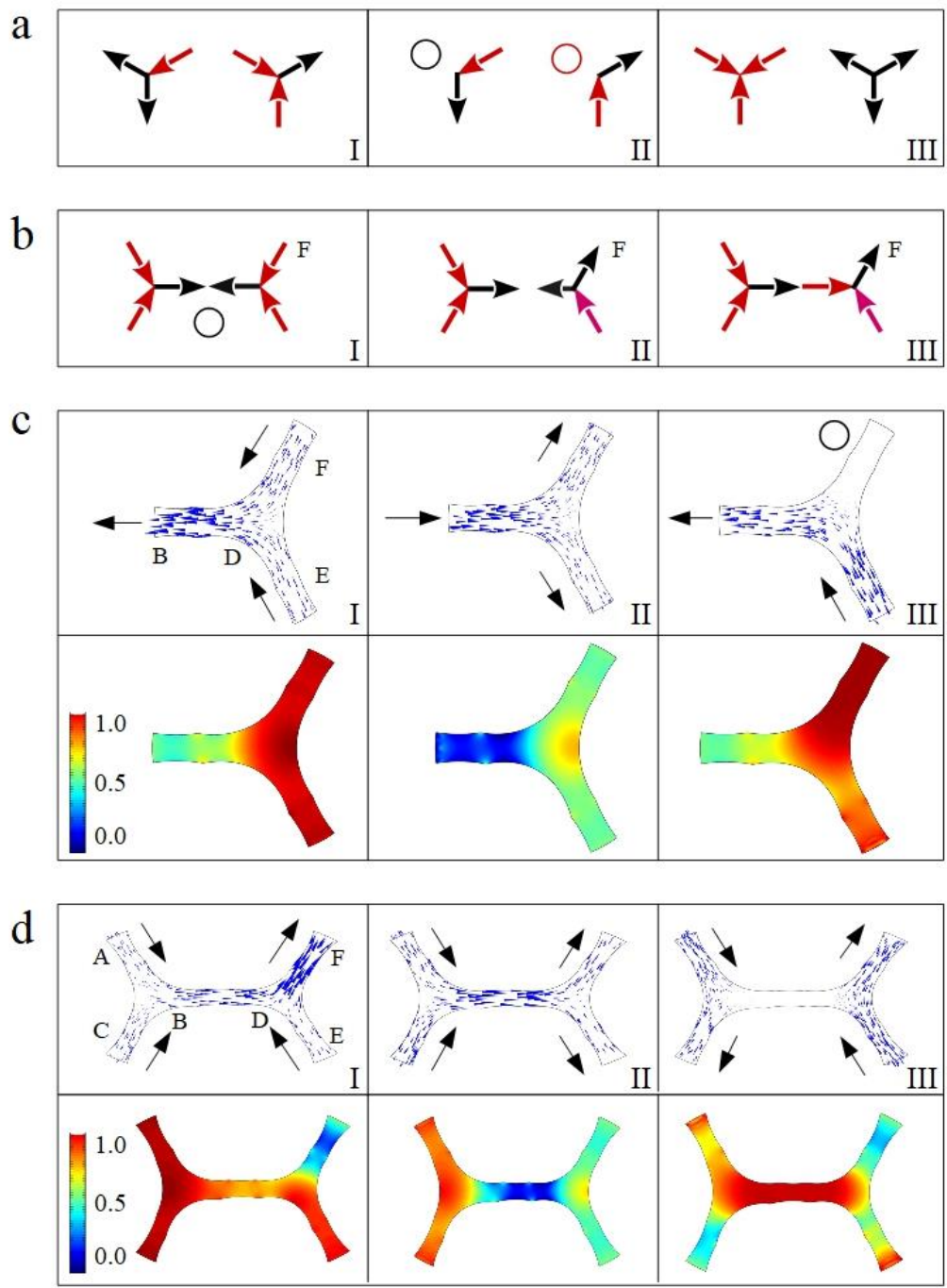

Figure 5: The combinations of all possible flow vectors at the Y-node and H-node, and its fluid dynamic simulation. (a) I) The two-in-one-out and one-in-two out configurations obey ice-rule. II) The one-in-one-out pattern. III) The three-in and three-out at the Y-node violate the ice rule. (b) I) Two opposite current meet in the middle vein to cause a jammed channel. II) The inflow at F reversed its direction. III) The traffic jam is dissolved. (c) The simulated velocity field at the Y-node. (d) The simulated velocity field at the H-node.

As illustrated in Fig 5b-I, the inflow at entrance F reverses direction, displacing jammed particles from the central vein (Fig 5b-II) and ultimately resolving the traffic congestion (Fig 5b-III). Throughout this process, both Y-shaped nodes maintain flow configurations that obey the ice rule. consequently, periodic flow reversal constitutes the key mechanism enabling the *Physarum* network to prevent traffic congestion, which would otherwise severely impair the efficiency of nutrient and signal transport. Notably, intersection nodes with more than four branch veins are exceedingly rare; nearly all nodes are Y-shaped. Furthermore, despite the existence of loops within the network, circulating flows around such loops were not observed.

Fig 5c presents the numerically simulated velocity vector field and pressure distribution at a Y-shaped node, obtained using the computational fluid dynamics (CFD) module of COMSOL-Multiphysics. In the two-in-one-out configuration (Fig 5c-I), the two inflow entrances (F and E) exhibit higher pressure than the outflow exit (B). Conversely, in the one-in-two-out configuration (Fig 5c-II), the inflow entrance(B) shows lower pressure than the two outflow exits (F and E). As shown in Fig 5c-III, the blocked entrance F attains the highest pressure, which drives flow from the entrance E toward exit B.

The flow field distribution across the H-shaped nodes is governed by the interaction between the two constituent Y-shaped nodes. As shown in Fig 5d-I, the high-pressure region drives three inflows from entrances, A, C and E toward the low-pressure exit F, with most particles concentrated along the exit channel DF. In contrast, for the configuration in Fig 5d-II where the two inflow entrances (A and C) are located on the same side of the central channel, particles primarily accumulate along this central channel, as its two ends are obstructed by higher pressure. In Fig 5d-III, where both inflow entrances and outflow exits are positioned on opposite sides of the central channel, the high pressure along the channel forces all particles out, resulting in a traffic congestion within the central channel. However, this jammed state persists for no more than 10 seconds before being resolved by a delayed reversal of flows in the branch veins connected to the H-node. These simulation finding are consistent with our experimental observations.

## 3 Discussion

The *Physarum* network is an active, adaptive system that expands toward nutrient sources and retracts from nutrient poor regions. Unlike unidirectional flows, the cytoplasmic flow within this network oscillates back and forth along the transport channels. This oscillatory flow, termed "venous shuttle flow", governs the network's expansion and retraction dynamics. In this study, we investigated how the venous shuttle flow redistributes at the network intersections. The measured flows at Y-shaped node satisfy Kirchhoff's current laws, thereby facilitating the redistribution of protoplasm throughout the entire network. The oscillations are not globally synchronous; instead, a phase difference exists among the branch veins converging at each node. This phase difference establishes a self-regulating flow mechanism that enhances the network's adaptive capacity. Furthermore, the venous shuttle flow prevents prolonged vein congestion, offering novel insights for addressing traffic congestion problems. Notably, the venous shuttle flow provides a biological analog for the spin ice model. In fact, the flow vectors consistently conform to the "ice rules", presenting a natural, intelligent solution to the frustrated configurations in spin ice systems. Beyond physics, the expansion and retraction strategies of *Physarum* network inspire the design of responsive biomaterials and intelligent robots for optimal transport applications.

## 4 Acknowledgment

Project supported by National Natural Science foundation of China (22193033).